\documentclass[11pt]{article}
\include{GrandMacros}

\usepackage{graphicx,verbatim,array,multicol, palatino}
\usepackage{amsthm, amsmath, amssymb,  setspace}
\usepackage{epsfig, url,epstopdf}
\usepackage{xcolor,colortbl}
\usepackage{color}
\usepackage{mathtools}
\usepackage{indentfirst}
\usepackage{hhline}

\definecolor{red}{rgb}{1,0,0}
\definecolor{blue}{rgb}{0,0,1}
\definecolor{green}{rgb}{0,0.6,0.4}

\begin{document}

\title{Feature-Based Classification of Networks}

\author{Ian Barnett\footnote{These authors contributed equally} \footnote{Harvard University, Department of Biostatistics, Boston, 02115, USA}  \\ 
Nishant Malik\footnotemark[1] \footnote{Dartmouth College, Department of Mathematics, Hanover, 03755, USA}  \\
Marieke L. Kuijjer\footnote{Dana Farber Cancer Institute, Boston, 02115, USA} \footnote{Harvard T.H. Chan School of Public Health, Boston, 02115, USA} \\
Peter J. Mucha\footnote{University of North Carolina, Department of Mathematics, Chapel Hill, 27599, USA} \\
Jukka-Pekka Onnela\footnotemark[2]
}

\date{}

\maketitle

\newpage

Network representations of systems from various scientific and societal domains are neither completely random nor fully regular, but instead appear to contain recurring structural building blocks \cite{milo2002network}. These features tend to be shared by networks belonging to the same broad class, such as the class of social networks or the class of biological networks. At a finer scale of classification within each such class, networks describing more similar systems tend to have more similar features. This occurs presumably because networks representing similar purposes or constructions would be expected to be generated by a shared set of domain specific mechanisms, and it should therefore be possible to classify these networks into categories based on their features at various structural levels. Here we describe and demonstrate a new, hybrid approach that combines manual selection of features of potential interest with existing automated classification methods. In particular, selecting well-known and well-studied features that have been used throughout social network analysis and network science \cite{wasserman1994social,newman2010networks} and then classifying with methods such as random forests \cite{james2013introduction} that are of special utility in the presence of feature collinearity, we find that we achieve higher accuracy, in shorter computation time, with greater interpretability of the network classification results.

Past work in the area of network classification has primarily focused on distinguishing networks from different categories using two different broad classes of approaches. In the first approach, network classification is carried out by examining certain specific structural features and investigating whether networks belonging to the same category are similar across one or more dimensions as defined by these features  \cite{ralaivola2005graph,pennacchiotti2011machine,onnela2012taxonomies,richiardi2011classifying}. In other words, in this approach the investigator manually chooses the structural characteristics of interest and more or less manually (informally) determines the regions of the feature space that correspond to different classes. These methods are scalable to large networks and yield results that are easily interpreted in terms of the characteristics of interest, but in practice they tend to lead to suboptimal classification accuracy. In the second approach, network classification is done by using very flexible machine learning classifiers that, when presented with a network as an input, classify its category or class as an output \cite{yanardag2015deep, gartner2003graph,kashima2003marginalized,borgwardt2005shortest,horvath2004cyclic,ramon2003expressivity,shervashidze2009fast,kondor2009graphlet,thoma2010discriminative,niepert2016learning}. To somewhat oversimplify, the first approach relies on manual feature specification followed by manual selection of a classification system, whereas the second approach is its opposite, relying on automated feature detection followed by automated classification. While the latter approach can yield very accurate class predictions, its computational cost typically scales poorly and, perhaps more importantly, the potentially opaque nature of the methodology may make it difficult to interpret the obtained results.

This paper presents a third, hybrid approach to the network classification problem. We first specify network features of interest manually and then use existing automatic methods, such as random forests, to carry out the classification using these features. In other words, our approach uses manual feature selection followed by automated classification. This approach enables one to leverage domain specific knowledge to specify a much broader set of relevant features. These features might be based on some standard network characteristics, such as vertex degree, betweenness centrality, or motif counts, but they can also incorporate nodal attributes, such as the sex or age of a person in a social network. It is possible to incorporate even richer information, such as data related to the functional or dynamic state of the nodes and edges. For example, in the context of network epidemiology, the frequency with which a node changes state from susceptible to infected in a contact network in the course of a spreading process could be used as a predictor. Then, since the classifier detects the relative importance of the different network features to the prediction, the resulting organization of networks can be better understood intuitively.

Our approach to the classification problem is scalable and its results are easily interpretable. The approach also leads to remarkably high classification accuracy, as we demonstrate by discerning different days of the week in unipartite social communication networks, distinguishing between different tumor body sites in bipartite biological transcription factor-gene regulatory networks, and testing the methodology on a collection of network classification benchmarks. Importantly, it is not clear \emph{a priori} what the best features might be as they would be expected to depend strongly on the domain of the network. Further, not all classifiers are equally suited to the task. In particular, many network properties are related to one another and their collinearity can cause problems when they are used as predictors. This calls for a classifier that handles collinear predictors well, as we discuss below.

We studied three different types of networks. First, to demonstrate classification on social networks, we constructed daily communication networks using call detail records from the largest telecom operator ($57\%$ market share) in a European country for three quarters of the year 2014. In these networks, undirected edges are placed between any two individuals who communicated with one another either via phone calls or text messages on the given day. We use different features of network structure and properties of network nodes to classify the networks into days of the week and, more generally, into weekday (Monday through Friday) vs. weekend (Saturday and Sunday) networks. Given the natural weekly periodicity in human behavior, we would expect the structure of these networks to reflect changes in the day-by-day social activity and communication patterns. Second, to demonstrate classification on biological networks, we used regulatory networks from tumor cells from patients with either lung (lung adenocarcinoma), brain (glioblastoma multiforme), or ovary (ovarian serous cystadenocarcinoma) cancer. For each sample, we constructed a bipartite network of genes and transcription factors with edge weights corresponding to the strength of regulation between a transcription factor and a gene for 113 transcription factors and 10,903 genes \cite{kuijjer2015estimating}. Given that gene expression levels would be expected to differ by tumor site, we would expect the properties of the bipartite regulatory networks to vary from site to site. Third, we investigated a variety of network classification benchmarks, including internet-based ego-centric social networks constructed from forum discussion threads and acting networks constructed from the Internet Movie Database (IMDb). For these benchmarks, we used network features to classify the forum thread networks by their topic and the acting networks by the movie genre.

Within each of these families of networks, we performed classification by a two-step process (Figure \ref{fig:dia1}): in the first step, we select and calculate features that may be pertinent to the classification problem for each network; in the second step, we train and test a classifier built upon these features. Importantly, because features are first selected manually based on available information and then further refined by the classifier, the set of features used for accurate classification varies depending on the family of networks of interest. Despite this, there are some common network features that we use in each setting, including average degree, global clustering coefficient, degree assortativity, and network size. We characterized the social networks constructed from call detail records using available specialized features, such as the fraction of male-female edges, and the fraction of edges between people who reside in the same zip code. The biological networks include features for the complete gene-transcription factor bipartite networks as well as for unipartite projections onto gene-gene and transcription factor - transcription factor space. No additional specialized features were used to classify the internet-based social network benchmarks.

After selecting the network features, the second step requires choosing the appropriate method for classification. We used three popular classification techniques: $k$-means, $k$-nearest-neighbors (KNN), and random forests. All three approaches are spatial classifiers, meaning they divide the feature space into regions and all networks that fall in the same region are assigned to the same class. The $k$-means and KNN tend to work well when the feature space is nearly linearly separable by class type, but can have difficulty with strong feature collinearity and more complicated class boundaries. On the other hand, random forests use a combination of multiple rectangular regions which allow for more flexibility with feature space partitioning. For each approach, the classifier is trained on a subset of the networks, and the classification accuracy of the three approaches is tested using the set of remaining networks that were not used in the fitting process.

None of the classifiers we used had difficulty separating weekends from weekdays in the phone-based social networks, with all methods achieving greater than $95\%$ prediction accuracy (Figure \ref{fig:comb_soc1}). In the random forest classifier, the fraction of edges connecting individuals residing in the same zip code was the most important feature (in terms of the fraction of trees sampled relying on this feature), being 4.5 times more important than the average feature used in classification (Figure \ref{fig:comb_soc2}). On the weekends, there was a clear increase in the proportion of ties that connect people from the same zip code. The second most important feature was network size, reflecting the marked decrease in the number of phones used on the weekend as compared to weekdays. The average age difference over all network edges increased by approximately one year on weekends compared to weekdays, leading to this feature being third in importance. This near perfect  accuracy by different classifiers indicates that the features for weekdays and weekends are easily distinguished from one another. As seen in Figures \ref{fig:comb_soc1} and \ref{fig:comb_soc2}, we similarly classified these networks into 7 groups corresponding to the days of the week, with lower accuracy overall but still excellent identification of particular days, e.g.\ with Saturdays well distinguished from Sundays but less accuracy for the middle of the weekdays.

Distinguishing tumor types based on their regulatory networks proved to be a more difficult task. The random forest classifier had an overall prediction accuracy of $68\%$ compared to the $62\%$ prediction accuracy of the KNN classifier using the same set of features. We observe that KNN did not perform as well as random forest because the tumor types did not form linearly separable clusters in the feature space, whereas the random forest classifier was able to more flexibly partition the space. The most important feature in the random forest classification of the tumor samples was degree assortativity in the projected gene-gene unipartite network, at $1.6$ times as important as the average feature. However, in contrast to the phone-based social network where a subset of the features were clearly driving the results, in this case there was a more uniform contribution from all selected features. As seen in Figure \ref{fig:RFclassBIO}(a), one of the lung tumor tissue samples was classified alongside the ovarian tumor tissue samples. This demonstrates how classification can be used to identify outliers to be checked for potential mislabeling. 

Random forests similarly outperformed KNN in the benchmark classification problems, with an average prediction accuracy margin of $3\%$ across the six benchmarks in Table 1. Moreover, and remarkably, our hybrid approach of manually selecting features and using random forests to automatically select their importance also outperformed three recently developed and significantly more complicated and computationally-intensive approaches to graph classification, namely graph kernels \cite{kondor2002diffusion}, deep graph kernels \cite{yanardag2015deep}, and convolutional neural networks \cite{fukushima1980neocognitron} (results for each reported in Ref.~\cite{niepert2016learning}). Given the value of domain specific knowledge for selecting and interpreting prospective features of importance, and given that random forests are easily trained on large data sets and allow for easy interpretation of results, our hybrid approach combining manual specification of features followed by automated classification on the selected features, appears to have a significant advantage in terms of precision of classification, cost of computation, and ease of interpretation.

\section*{Acknowledgements}
We thank Kenth Engø-Monsen at Telenor Research for making the CDR-data available for this research. We thank Kimberly Glass and members of the Onnela lab for their feedback and useful discussion. We also acknowledge Nic Larsen, Natalie Stanley, and Sean Xiao for helping identify benchmarks and other network classification methods for comparison with our approach. MLK acknowledges support from the Charles A. King Trust Program and the National Cancer Institute Specialized Programs of Research Excellence. PJM acknowledges support from the James S. McDonnell Foundation 21st Century Science Initiative - Complex Systems Scholar Award grant $\#220020315$ and the Eunice Kennedy Shriver National Institute of Child Health \& Human Development of the National Institutes of Health under Award Number R01HD075712. The content is solely the responsibility of the authors and does not necessarily represent the official views of any funding agency.
\section*{Author contributions statement}
IB, NM, PJM and JP planned the analyses, IB and NM performed the analyses,  and MLK produced the regulatory networks. All authors together wrote the manuscript.

\section*{Additional information}

The authors declare no competing financial interests.

\bibliography{WeekEndDay}

\clearpage

\begin{figure}[ht]
\includegraphics[width=\columnwidth]{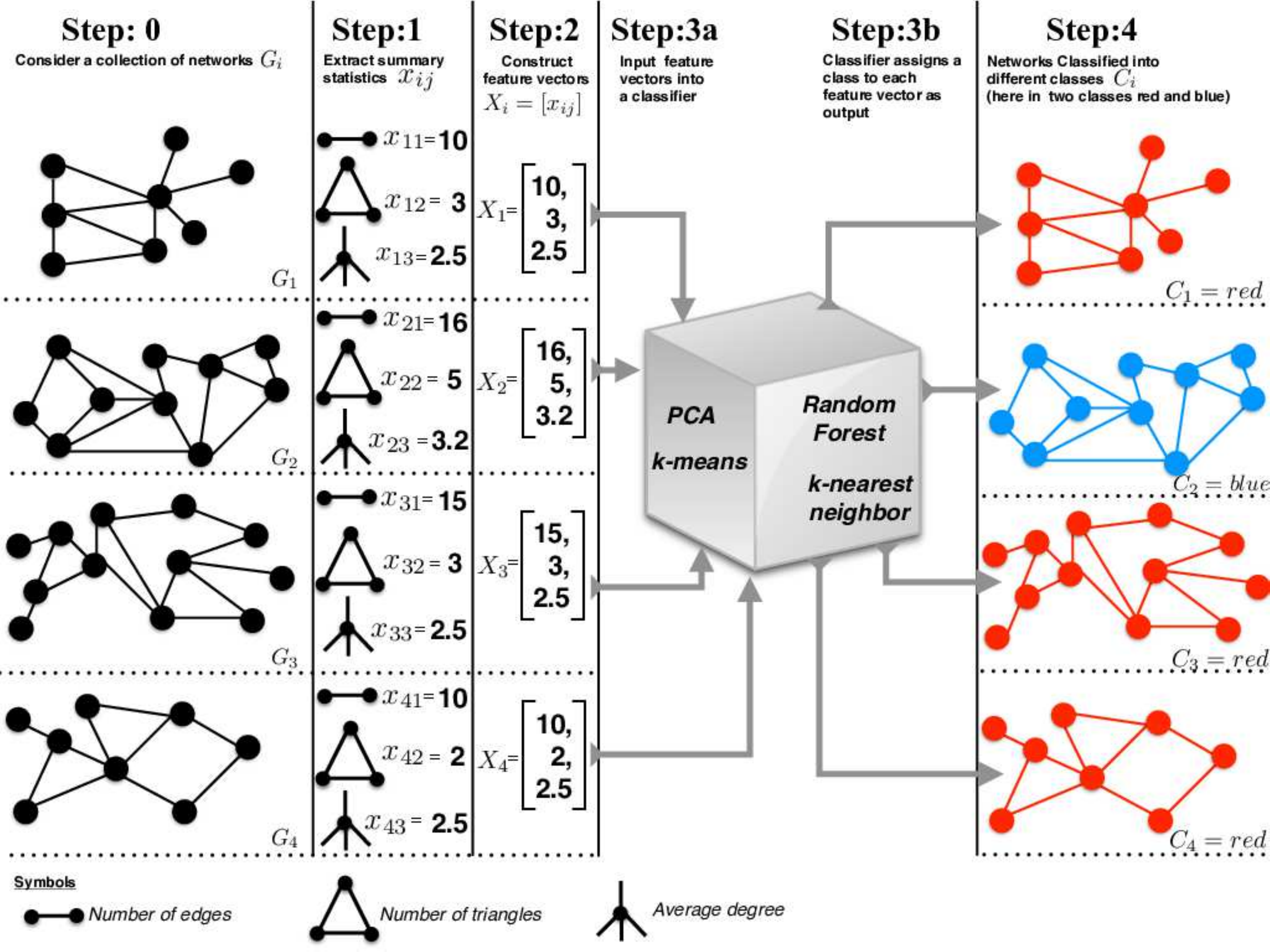}
\caption{\textbf{A schematic illustrating the steps in our network classification approach}. Here $\{G_i\}$ represents a collection of networks with known class labels, $x_{ij}$ is the $j$th feature of the $i$th network, and $X_i=[x_{ij}]^T$ is the (column) feature vector corresponding to the $i$th network. In principle, one could use several different classifiers, such as principal component analysis (PCA),  $k$-means clustering, $k$-nearest neighbor, and, important to our findings, random forests.}
\label{fig:dia1}
\end{figure}

\clearpage

\begin{figure}[ht]
\centerline{\includegraphics[width=.75\columnwidth]{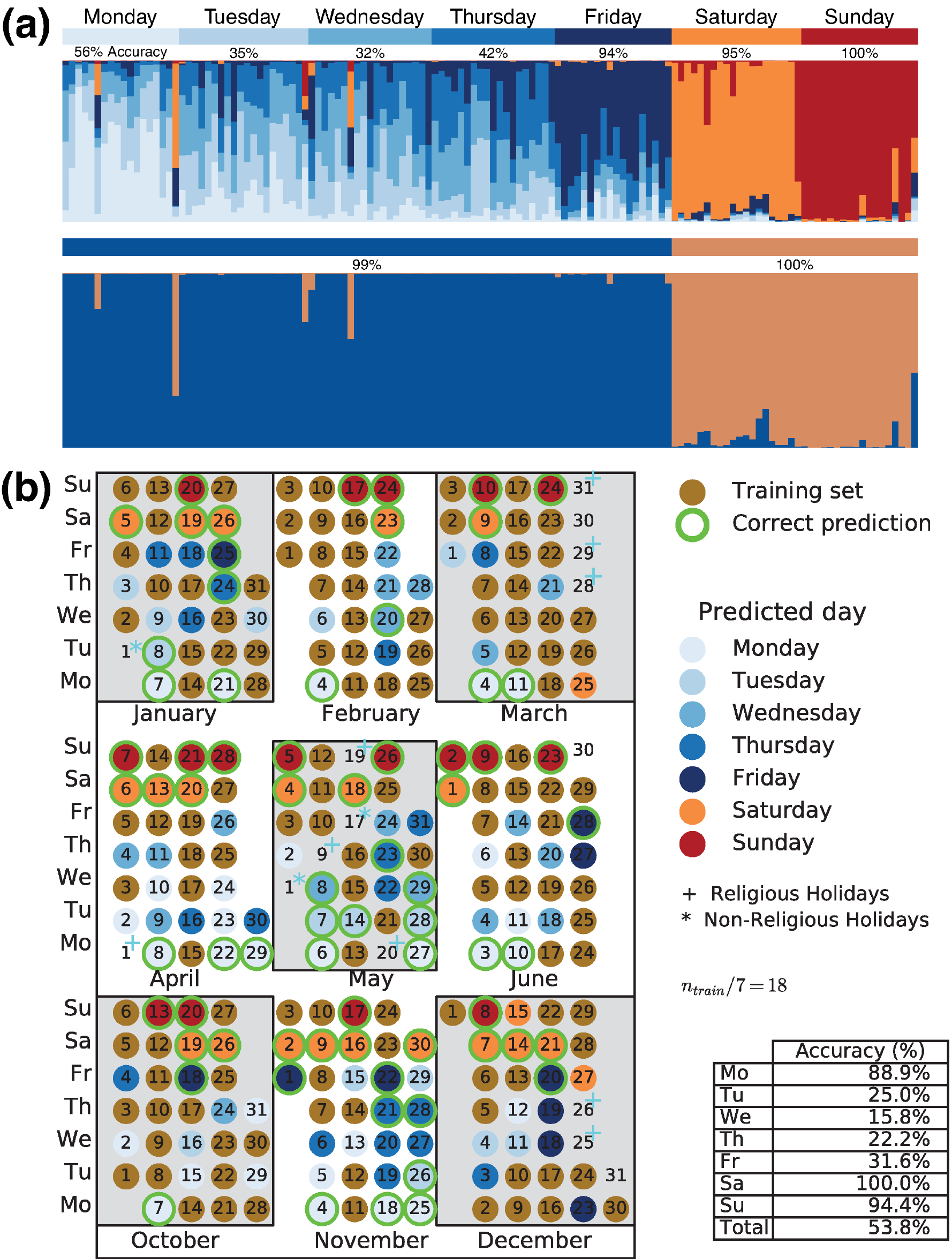}}
\vspace{1pc}
\caption{(a) \textbf{Random forest classification of days of the week}: Using odd-numbered days of the data set for training, the classification of each even day is displayed as a column. The performance of the 7-day classifier is displayed in the top row with the binary weekend/weekday classifier in the bottom row. Each column represents the color-coded probabilities of a day being classified as a day of the corresponding color. In the top row, a day is correctly classified if that day has the largest classification probability. For the bottom row, the larger of the two binary classification probabilities is used to guide the classification.  All nationally-recognized holidays were removed from both the training and testing datasets as they would be expected to have unusual social dynamics. (b) \textbf{KNN classification of days of the week}: This visualizes a single realization of classification of days of the week using KNN, where $n_{train}$ is the total number of days used for the training set, which included equal number of days of each day of the week.}
\label{fig:comb_soc1}
\end{figure}

\begin{figure}[ht]
\centerline{\includegraphics[width=1\columnwidth]{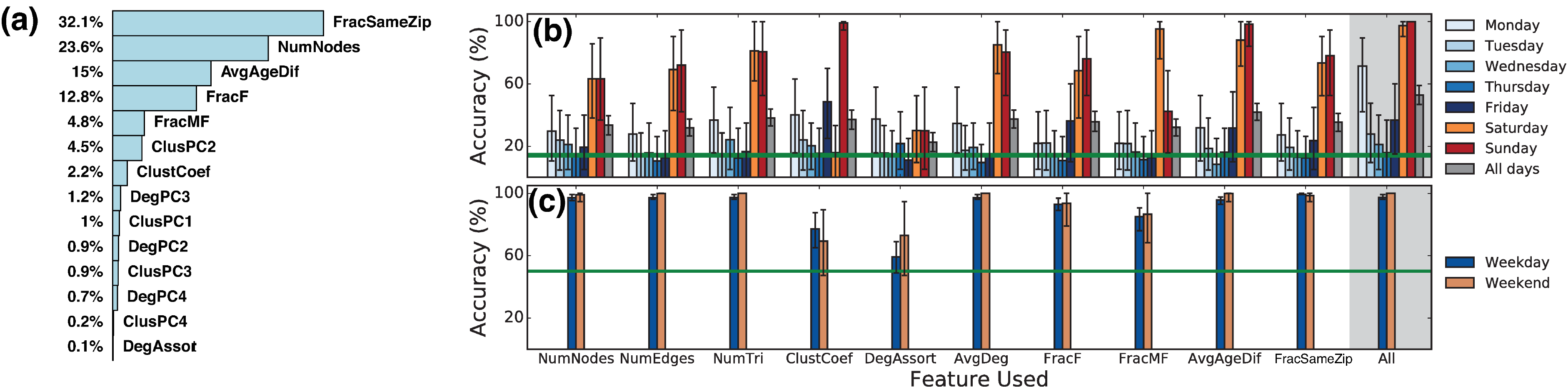}}
\vspace{1pc}
\caption{(a) \textbf{Feature importance in the weekend-weekday classification random forest}:  Feature importance is calculated from the mean decrease in tree leaf impurity over the full random forest as measured by the Gini index. Percentages are the decrease in impurity for each feature, scaled so they sum to 100\%. Three redundant features are not displayed due to their strong correlation with the NumNodes feature. Detailed descriptions of the variables are provided in the supplemental materials. (b) \textbf{KNN day-of-week classification}: Each set of bars represents accuracy over multiple realizations of the seven days of the week and the combined accuracy over seven days (grey bars), using a single indicated feature. The last block, highlighted in grey, represents accuracy using all selected features. The green line represents the null rate of classification. (c) \textbf{KNN weekend-weekday classification}: Bars represent accuracy of classifying weekdays and weekend days using the indicated features. The green line again represents the null rate.} 
\label{fig:comb_soc2}
\end{figure}

\clearpage
\begin{figure}[ht]
\centerline{\includegraphics[width=.9\columnwidth]{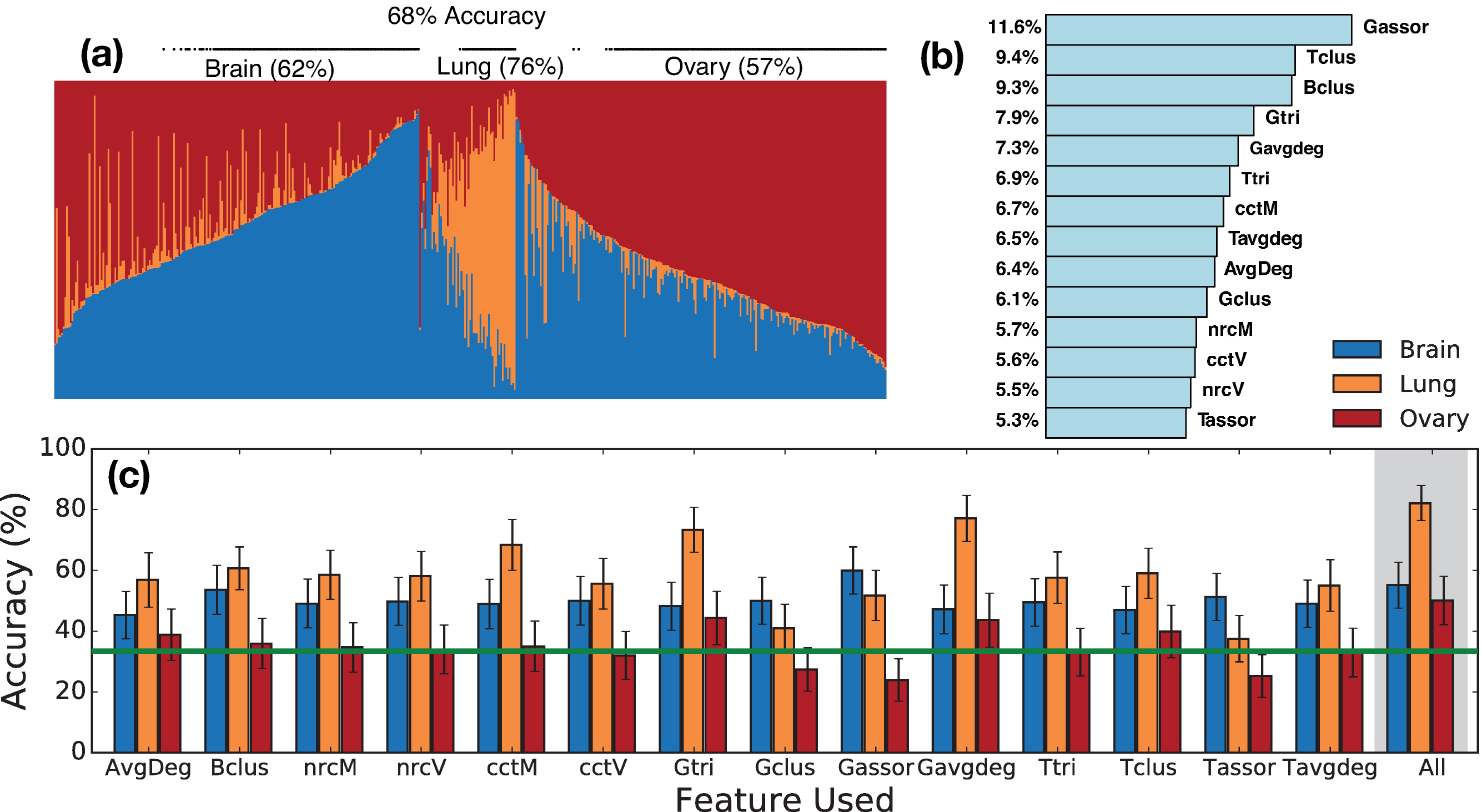}}
\vspace{1pc}
\caption{(a) \textbf{Random forest classification of cancer types}: Each of the $483$ columns represents random forest classifier probabilities as stacked bars for a tissue sample in the test set, with blue, orange and red bars representing probabilities assigned to brain, lung, and ovary cancers, respectively. Each sample is then classified by the largest of these three probabilities, and correct classification is indicated by a black dot above the corresponding column. Overall, $68\%$ of tissue samples were correctly classified ($32\%$ misclassified). (b) \textbf{Feature importance in the tumor type classification random forest}: Feature importance is calculated from the mean decrease in tree leaf impurity over the full random forest as measured by the Gini index. Percentages are the decrease in impurity for each feature, scaled so they sum to 100\%. Detailed descriptions of the variables are provided in the supplemental materials. (c) \textbf{KNN classification of cancer types}:  Each set of bars represents accuracy of the three types, using a single indicated feature. The last block, highlighted in grey, represents accuracy using all selected features. The green line represents the null rate of classification. Overall, $62\%$ of tissue samples were correctly classified using this method. Error bars indicate the standard deviation of mean accuracy over $10,000$ realizations.}
\label{fig:RFclassBIO}
\end{figure}

\clearpage

\begin{table}[]
\centering
\begin{tabular}{|l|l|l|l|l|l|}
\hline
Dataset            & RF         & KNN         & GK\cite{}          & DGK\cite{}         & PSCN\cite{}        \\
\hline
COLLAB             & $76.5\pm1.68$ & $72.69\pm0.80$ & $72.84\pm0.28$ & $73.09\pm0.25$ & $72.60\pm2.15$  \\
\hline
IMDB-BINARY       & $72.4\pm4.69$ & $37.03\pm1.90$  & $65.87\pm0.98$ & $66.96\pm0.56$ & $71.00\pm2.29$    \\
\hline
IMDB-MULTI        & $47.8\pm3.55$ & $42.40\pm2.70$  & $43.89\pm0.38$ & $44.55\pm0.52$ & $45.23\pm2.84$ \\
\hline
REDDIT-BINARY     & $88.7\pm1.99$ & $87.63\pm0.82$ & $77.34\pm0.18$ & $78.04\pm0.39$ & $86.30\pm1.58$  \\
\hline
REDDIT-MULTI-5K  & $50.9\pm2.07$ & $49.04\pm0.77$ & $41.01\pm0.17$ & $41.27\pm0.18$ & $49.10\pm0.70$   \\
\hline
REDDIT-MULTI-12K & $42.7\pm1.28$ & $38.21\pm0.49$ & $31.82\pm0.08$ & $32.22\pm0.10$ & $41.32\pm0.42$ \\
\hline
\end{tabular}
\caption{\textbf{Classification accuracy for benchmark social network data sets.} Results expressed as $\%$, from 10-fold cross-validation to obtain out-of-sample accuracy estimates and their standard deviations, for the random forests (RF) and KNN classifiers used here, compared to results for Graph Kernels (GK), Deep Graph Kernels (DGK), and convolutional neural networks (PSCN), as reported in Ref.~\cite{niepert2016learning}. }
\label{tab:socnetclass}
\end{table}

\clearpage

\section{Methods}

\subsection{Network feature extraction}

Feature-based classification is a two-step procedure, regardless of the application. First, a set of contextually important network features are selected to be calculated for each network. Second, the data is split into training and testing datasets and the features are fed into the classifier of choice. For a more detailed schematic see Figure \ref{fig:dia1}.

\subsubsection{Call activity social networks}

For the social network setting we use the call detail record (CDR) data from the first, second, and fourth quarter of the year 2014 from a European country's leading telecom operator, which had $57\%$ market share. We denote $N$ as the number of days in the training set. Let $X_i = [x_{i1},\dots,x_{ip}]^T$ be the $p$ features of the $i$th day. National holidays were removed from the analysis because of likely anomolous social behavior on those days. 

For each day, the daily call network is constructed by assigning an edge between any two individuals who are in contact by phone on that day. For each day's network, a variety of network features are extracted: the network size (excluding all nodes with degree 0), average clustering coefficient, degree assortativity, fraction of nodes that are female, fraction of edges that are male-female, average age difference over all edge pairs, the fraction of edge pairs from the same zip code, the first four principal components from the degree distribution, and the first four principal components from the clustering coefficient distribution. These features are then used in the selected classifers to predict whether or not a social call network corresponds to a weekend or a weekday, or in the 7-day classifier to a specific day of the week.

\subsubsection{Biological networks}

Tumor gene expression data was downloaded from The Cancer Genome Atlas for 1217 patients with cancer of the lung (lung adenocarcinoma), brain (glioblastoma multiforme), or ovary (ovarian serous cystadenocarcinoma). For each sample we reconstructed a bipartite network with edge weights corresponding to the strength of regulation between a transcription factor and a gene, across 10,903 genes and 113 transcription factors \cite{kuijjer2015estimating}. In this setting, $N = 547$ is the number of individuals in the training set and $X_i = [x_{i1},\dots,x_{ip}]^T$ represent the $p$ features of the $i$th sample, or individual.

For simplicity, we threshold edge weights in the bipartite network. For each edge, only the top $q\%$ of edge weights across all 1217 networks are declared to be edges, where $q \in [0,100]$ is the chosen threshold. In other words, for each possible edge, only the networks with edge weights in the top $1217 *(1-q)$ will contain that edge. A large $q$ leads to sparse networks whereas small $q$ leads to dense networks. We use $q=95$ but to test the sensitivity to $q$ we repeat the analysis using a variety of thresholds.

After the bipartite networks are constructed for each sample, each network is projected onto two unipartite sets, giving gene-gene networks and transcription factor - transcription factor networks. Selected features are then extracted from all three network representations. On the bipartite networks, we use average degree, average bipartite clustering coefficient, the mean and variance of node redundancy, and the mean and variance of node closeness centrality. On the unipartite projections, we use the average degree, the number of triangles, average clustering coefficient, and degree assortativity. These features are used to predict what type of cancer tumor the sample was taken from in the remaining 546 samples in the testing set.

\subsubsection{Benchmark online and acting social networks}

We also compare our approach to six benchmark social network classification tasks previously considered in the literature \cite{niepert2016learning}.

The online forum Reddit contains many discussion threads about assorted topics. Social networks were constructed for each thread by considering users as nodes and by placing an undirected edge between two users when one had responded to the other in that thread. Some subreddits have more specialized topics. The REDDIT-BINARY dataset is used to classify threads as either belonging to a question/answer-based subreddit or a discussion-based subreddit. The REDDIT-MULTI-5K dataset contains 5,000 thread networks across five different subreddits and the REDDIT-MULTI-12K dataset contains 12,000 thread networks from eleven different subreddits. In both data sets, the aim is to classify a thread into its correct subreddit.

COLLAB is a scientific-collaboration data set, where ego-based networks of researchers from three different fields are constructed with edges to other researchers the ego has collaborated with. The goal is to classify these ego-networks into their correct field. 

The IMDB-BINARY dataset constructs ego-based networks around every actor where edges are formed between actors that appear in the same movie together. Networks are constructed for two genres, \textit{Action} and \textit{Romance} (ignoring any movie in the union of the two), with the aim of classifying each ego-network into the correct genre. The IMDB-MULTI dataset is similar, but considers three genres, \textit{Comedy}, \textit{Romance}, and \textit{Sci-Fi}.

These six benchmark data sets were previously used to test the classification performance of two graph kernel approaches, Graph Kernels (GK) and Deep Graph Kernels (DGK)\cite{yanardag2015deep}, and an approach using convolutional neural networks (PSCN)\cite{niepert2016learning}. We compare our approach with the GK, DGK and PSCN results reported in Ref.~\cite{niepert2016learning}. For each network, we extracted six features to use in our classification: number of nodes, number of edges, average degree, degree assortativity, number of triangles, and the global clustering coefficient. Following the reporting of results in Ref.~\cite{niepert2016learning}, the accuracy of each of our classifiers was evaluated using 10-fold cross validation.

\subsection{Data driven network classification}

\subsubsection{Spatial classifiers: KNN and K-means}

For the $KNN$ classifier, we start with a training set $\mathcal{T}$ containing the feature vectors  ${{X}_i} = [x_{i1},\dots,x_{ip}]^T$, where $p$ is the number of features in the $i$th sample. Each feature vector  ${{X}}_i$ in $\mathcal{T}$ is preassigned a known class ${{Y}}_i \in \{1,\dots,c \}$. These classes could be days of the week as in CDR data set or disease sites as in the cancer data set. We find the $k$-nearest neighbors of a new feature vector ${{X}}_j$ in the prediction set  $\mathcal{P}$ using Euclidean distance $d({{X}}_i,X_j)=\sqrt{ \sum_{l=1}^{l=p} (x_{il}-x_{jl})^2 }$ and classify it into the ${{Y}}_j$ class, ${{Y}}_j \in \{1,\dots,c \}$ by the majority vote among the $k$-nearest neighbors. 

In contrast, $k$-means clustering provides an unsupervised classification system, wherein one partitions the complete set of feature vectors  $\{{{X}_i} \}_{i=1}^{N}$ into a set $C=\{C_1,C_2, \dots, C_k\}$ of $k \leq N$ clusters.  
These clusters are found by minimizing the square of the distance from the data points ${{{X}_j}} $ to the center of a cluster  i.e., solving  $$\underset{C_i}{\mathrm{arg min}} \displaystyle \sum_{i=1}^{k} ~~ \sum_{X_j \in C_i }( \lVert  { X_j-\mu_i} \rVert)^2, $$
where $\mu_i, ~ i=1,\dots,k$ is the position of cluster $C_i$. After this clustering, available known classification properties within each cluster can be assessed by various measures.

We refer the reader to Ref.~\cite{friedman2001elements} for a far more complete description of different methods for classifying and assessing classifications.

\subsubsection{Random forest classifier}
With $p$ features extracted, classification trees \cite{friedman2001elements} can be used to identify the subset of features that are important in distinguishing classes of networks (i.e. weekend days from weekdays). To begin construction of a tree, the data is split into the two groups that best separate the classes. Specifically, let 
\begin{equation}
R_1(j,s)=\{X_i : x_{ij}<s \} \mbox{ and } R_2(j,s)=\{X_i : x_{ij}\geq s \}
\end{equation}
be the regions that separate the data into two groups. Consider, for example, the classification of CDR social networks into weekend days and weekdays. Letting $\hat{p}_k$ be the fraction of data points in region $R_k$ that are weekdays, the regions that best separate the weekend days from the weekdays are determined by minimizing $\hat{p}_{1}(1-\hat{p}_{1}) + \hat{p}_{2}(1-\hat{p}_{2})$
with respect to $j$ and $s$. The $\hat{p}_{k}(1-\hat{p}_{k})$ function here is known as the \textit{Gini index}, penalizing the $k$th region if $\hat{p}_{k}$ is far from $0$ or $1$, as this indicates that the region does not separate the weekdays from weekend days very well. The minimization of the \textit{Gini index} only considers each individual branch of the classification tree at a time.

This process is repeated on the two resulting branches. This is repeated further until the data has been split too many times and there is only one data point in one of the branches, at which point the splitting on that branch terminates. For the minimization occuring at each branching, the random forest approach is as described above except we only consider a random subset $m \leq p$ of the features (we use $m=4 \approx \sqrt{p}$) while also bootstrapping the data at each branching. This introduces randomness into the tree building process, and $B=10,000$ such random classification trees are built. In order to classify a new data point, $x$, let $\hat{C}_b(x)$ be the class prediction of the $b$th random forest tree. The classification of $x$ is determined to be the majority vote over all $\{\hat{C}_b(x)\}_1^B$.

To apply this procedure to more than two outcomes (seven day classification as opposed to weekend versus weekday), the procedure is similar except the Gini index becomes $\sum_{k=1}^7\hat{p}_{rk}(1-\hat{p}_{rk})$ where $\hat{p}_{rk}$ represents the proportion of data points in region $r$ that represent day of the week $k$. The classification tree is built using odd days over all three available quarters, and the model is tested on the even days.

This same approach can applied to the biological network context to predict tumor type. For tumor type we consider three possible outcomes: brain, lung, and ovary. In our implementation, $p$ is the same so again we use $m=4 \approx \sqrt{p}$.

\newpage

\section{Supplemental Materials}

\subsection*{Further exploration of CDR data set}

\begin{figure}
\centerline{\includegraphics[width=0.7\columnwidth]{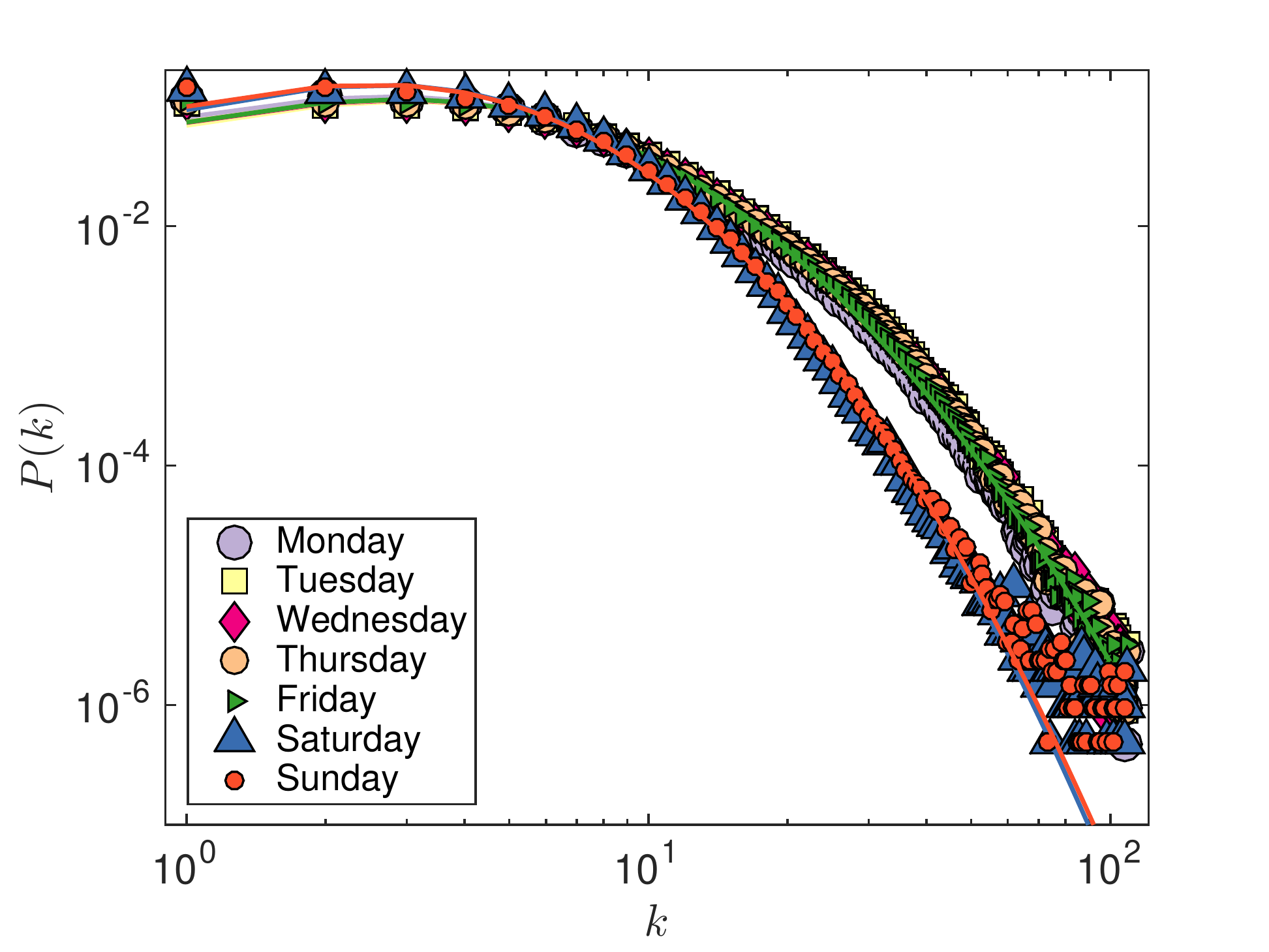}}
\vspace{1pc}
\caption{\textbf{Degree distribution for each day of the week.} Thick lines are fitted distributions given by Eq.~\ref{eq:dist1}, with values of the fitted parameters given in Table~\ref{tab:tab1}.} 
\label{fig:degdist}
\end{figure}

To further identify differences between the days of the week, we have estimated the degree distributions $p(k)$ for individual days, fitting them to lognormal distributions:  
\begin{equation}
p(k)= \frac{1}{k\sigma \sqrt{2\pi}}e^{\frac{-(\ln k-\mu)^2}{2\sigma^2} }.
\label{eq:dist1}
\end{equation}
Table~\ref{tab:tab1} provides the values of the fitted parameters and their estimated confidence intervals, whereas in Fig.~\ref{fig:degdist} we have plotted the empirical and fitted degree distributions. In particular, we observe in Fig.~\ref{fig:degdist} that weekends appear to have distinct distributions from the weekdays. The parameters of the fitted distributions are similarly distinct for weekdays compared to weekends (see Table~\ref{tab:tab1}).

\begin{figure}
\centerline{\includegraphics[width=0.7\columnwidth]{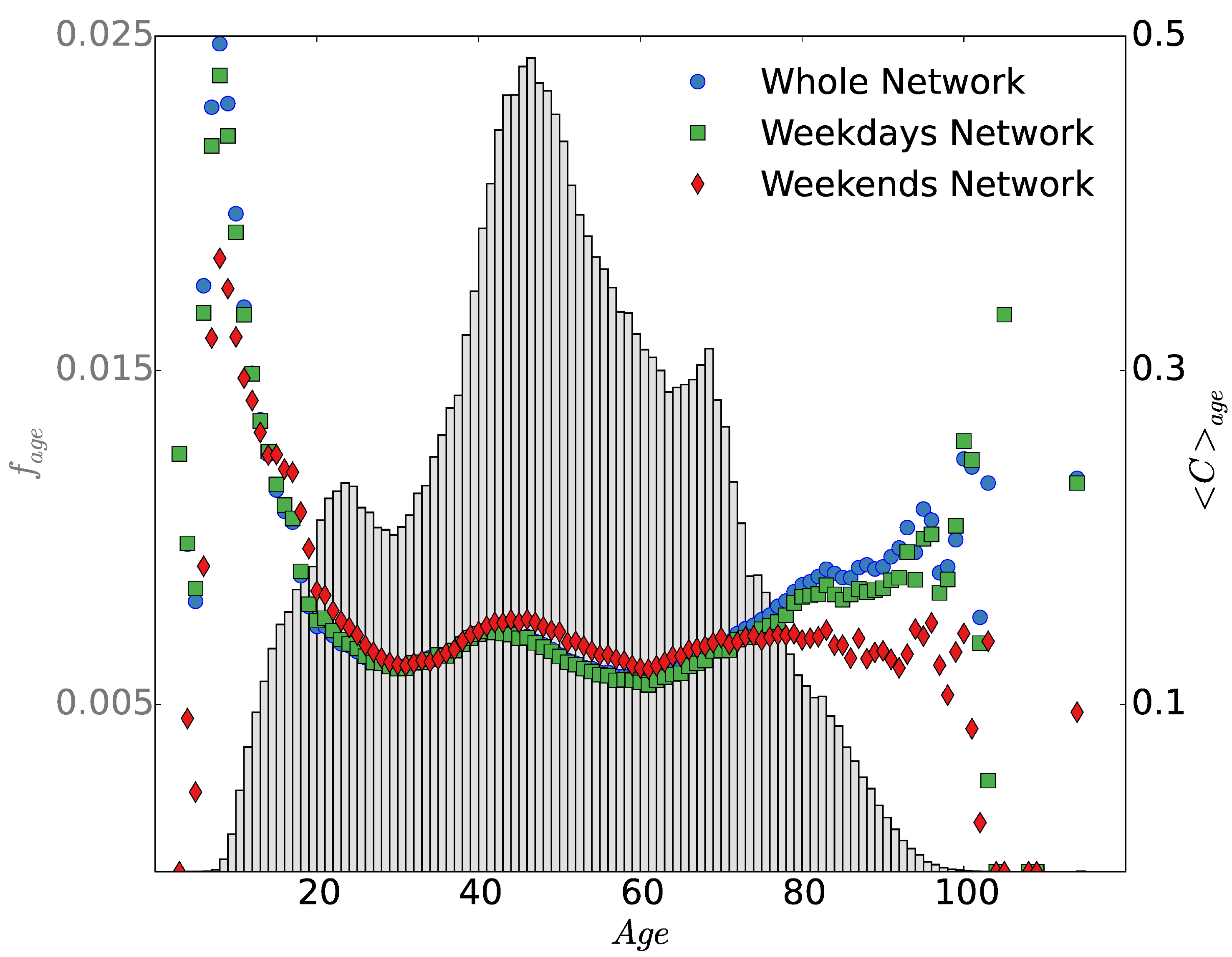}}
\vspace{1pc}
\caption{\textbf{Distribution of age and average clustering.} Here we consider three networks, one built from the whole data set across the three quarters of the year (whole network), one built across the data set but limited to week days (weekdays network), and finally one built across the data set but limited to weekends (weekends network). The grey histogram gives the distribution of age in the CDR data set.  Blue dots  are the average local clustering coefficients for nodes with the given age for the whole network data, whereas green squares and red diamonds represent the average clustering coefficients for the weekday and weekend networks, respectively. } 
\label{fig:ageclusdist}
\end{figure}

\begin{figure}
\centerline{\includegraphics[scale=.6]{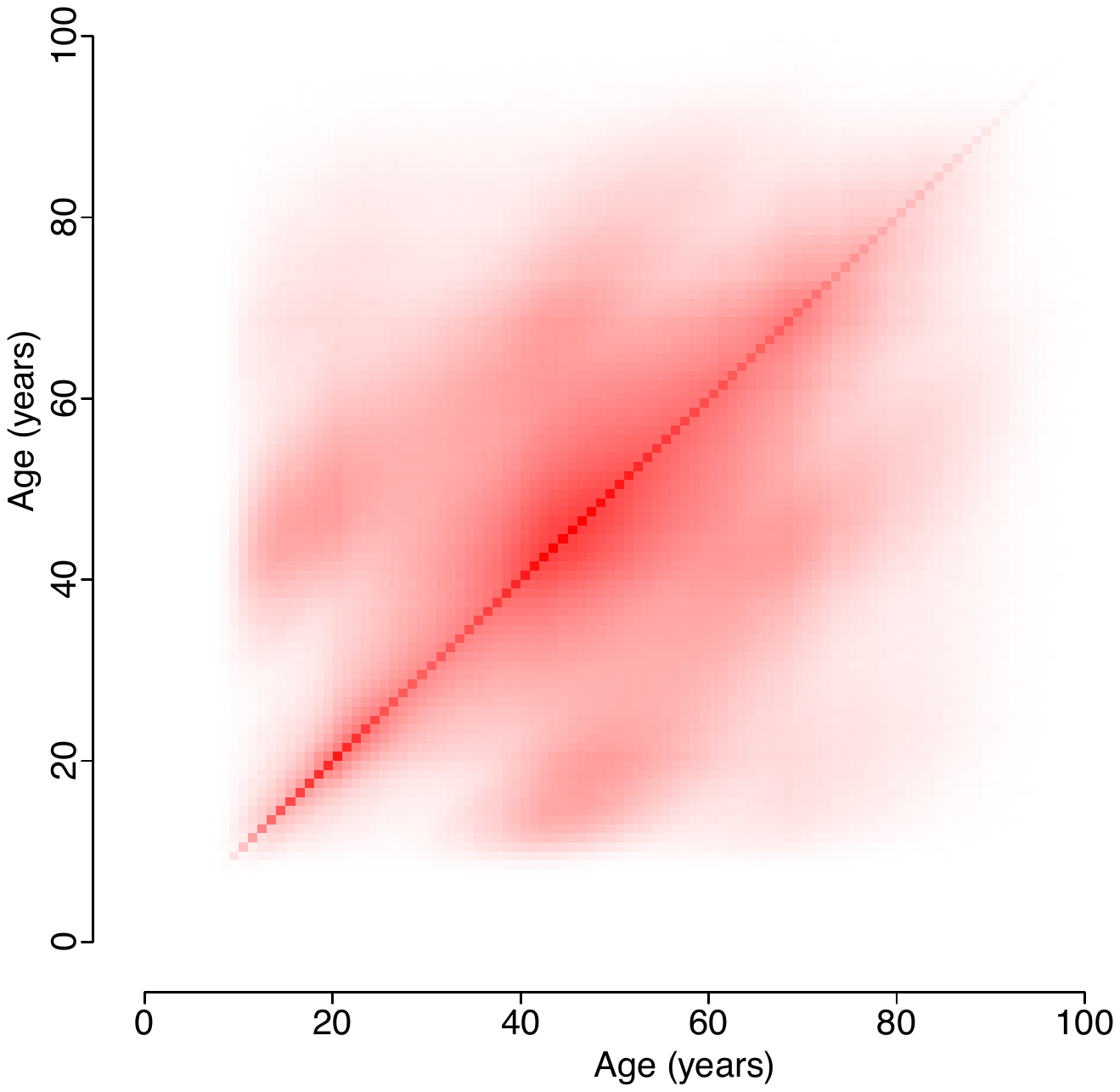}}
\vspace{1pc}
\caption{\textbf{Frequency of communication between age groups.} The full social network from the combined Q1, Q2, and Q4 of 2013 is used to count the number of age-age network edges. High frequency age-age connections are dark red, with less common age-age pairings in light red. The dark diagonal corresponds to communications within the same age group. The red intensity of the $(i,j)$ cell is $(x_{ij}/\max_{ij}{x_{ij}} )^4$ where $x_{ij}$ is the total number of edges people of age $i$ share with people of age $j$.} 
\label{fig:AgeByAge}
\end{figure}

Age plays a significant role in the way people use mobile phones, hence we might expect that communication patterns are influenced by user age. Average age difference across communication ties emerged as an important feature for the CDR data set (see Fig. 4). In Fig.~\ref{fig:ageclusdist}, we plot the distribution of age in the data and the corresponding average clustering in the network corresponding to a particular age group. One of the striking features here is that weekend and weekday networks show distinct patterns for average clustering versus age.  We also observe rather higher clustering for ages below 20, probably implying that these users interact within small tightly knit local network neighborhoods. In addition, most communication occurs between individuals of similar age, while there also appears to be a generational gap in high frequency communication between people approximately 25 years apart (see Fig.~\ref{fig:AgeByAge}), which likely reflects parent-child communication.

\begin{table}
\begin{center}
\begin{tabular}{ l | c | c | c}
\hline 
         Day    &         $\mu$     &   $\sigma$  & p-value \\
 \hline  
 Monday  & 1.794 $\pm$ 0.004 &  0.693 $\pm$ 0.003 & 0.656 \\ 
 Tuesday & 1.885 $\pm$ 0.004 &  0.710 $\pm$ 0.003 & 0.449 \\
  Wednesday & 1.860 $\pm$ 0.004 &  0.710 $\pm$ 0.003 & 0.989  \\ 
  Thursday & 1.853 $\pm$ 0.004 &  0.708 $\pm$ 0.003 & 0.811 \\ 
   Friday & 1.851 $\pm$ 0.004 &  0.702 $\pm$ 0.003 & 0.709 \\ 
   Saturday & 1.650 $\pm$ 0.004 &  0.603 $\pm$ 0.003 & 0.360  \\ 
   Sunday & 1.636 $\pm$ 0.004 &  0.613 $\pm$ 0.003 & 0.498 \\
\hline
\end{tabular}
\end{center}
\caption{\textbf{Values of fitted log-normal parameters and $95 \%$ confidence intervals.}  The confidence intervals were constructed assuming the asymptotic normality of the maximum likelihood estimate.  The p-values are obtained employing two-sample Kolmogorov-Smirnov tests under the null hypothesis that the fitted distribution and the sample distribution are the same continuous distribution. The test indicates that the fitted distribution and the sample distribution are the same.} 
\label{tab:tab1}
\end{table}

\subsection*{Classification of weekend days and weekdays in CDR data}

\begin{figure}
\centerline{\includegraphics[width=0.875\columnwidth]{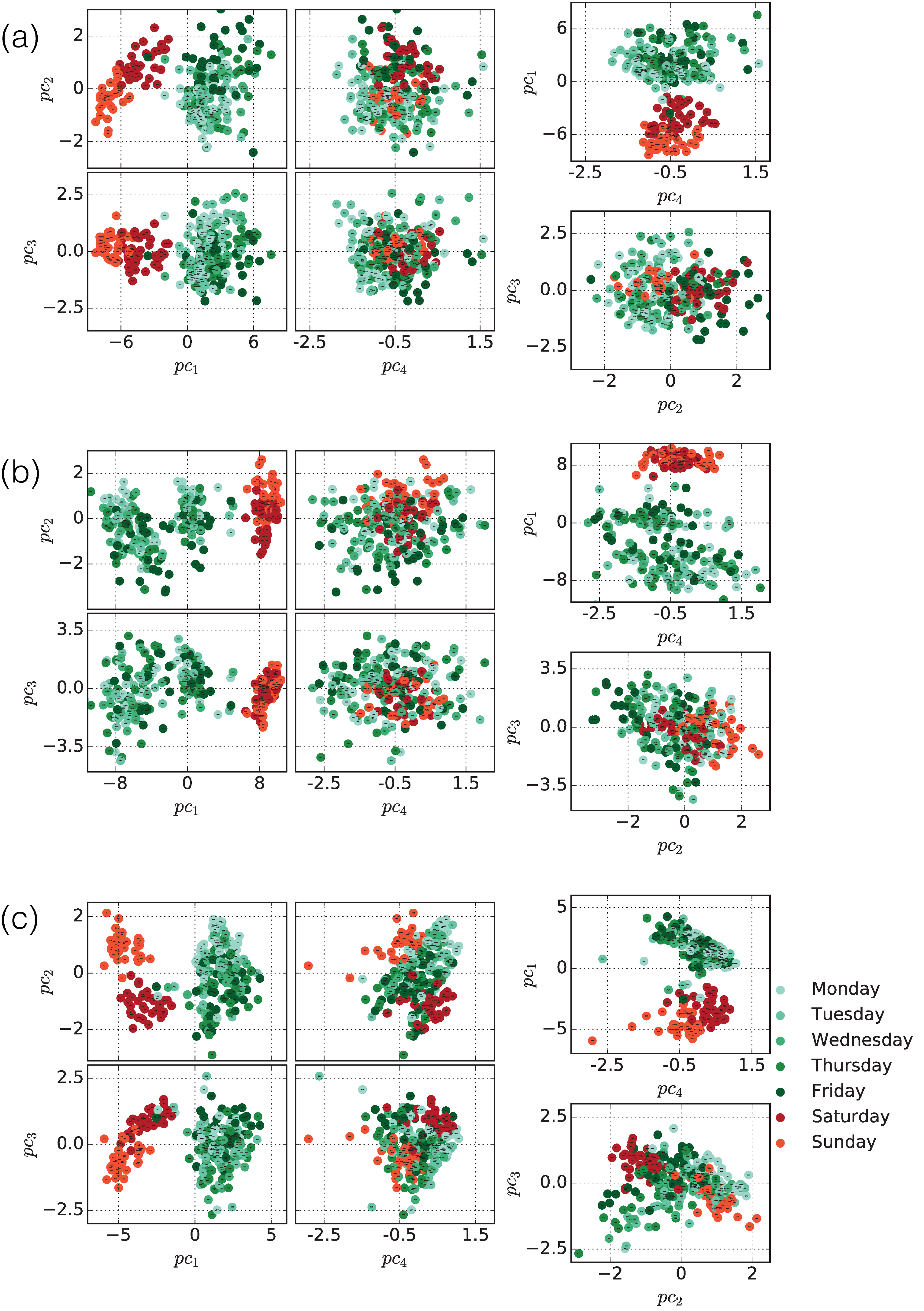}}
\vspace{1pc}
\caption{\textbf{PCA based classification of CDR data set.} (a) The first four PCA components for local clustering coefficient distributions of daily networks. (b)  The first four PCA components for degree distributions of daily networks. (c) The first four PCA components for feature vectors constructed from features listed in Table~\ref{tab:codeWK}.}
\label{fig:degdistPCA}
\end{figure}

We also employed PCA analysis and k-means clustering to classify weekend days and weekdays. In Fig.~\ref{fig:degdistPCA}, PCA was performed on three different feature vectors: degree distributions, distributions of local clustering coefficients, and the list of features in Table~\ref{tab:codeWK}. In each case, the first two PCA components were able to discriminate between weekend days and weekdays. 

\begin{figure}
\centerline{\includegraphics[width=1.0\columnwidth]{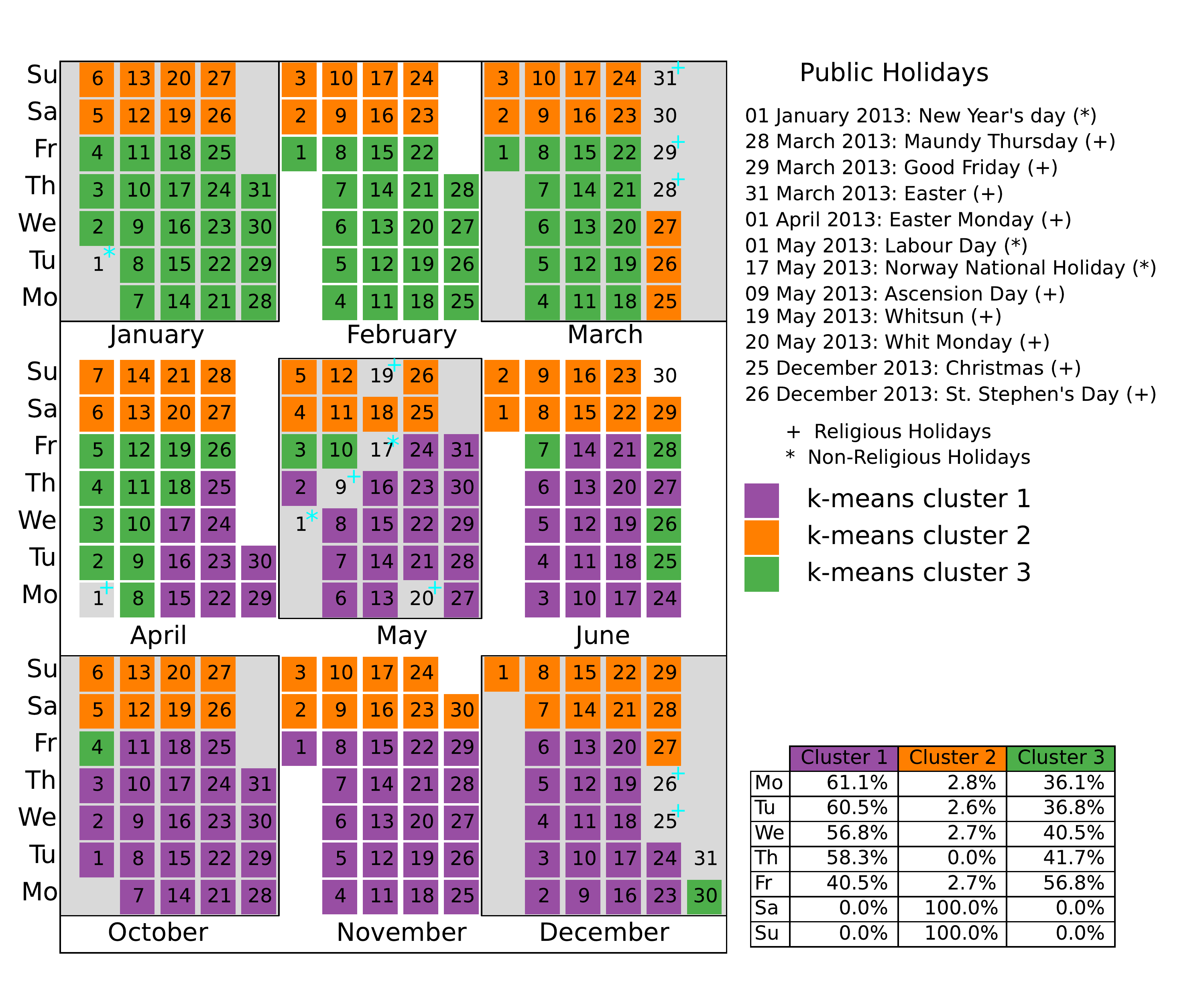}}
\vspace{1pc}
\caption{\textbf{Classification of days from daily call record data using k-means clustering.} Here we used degree distribution of the extracted daily networks as the feature vector. Holidays were removed from the data prior to running the classification routine.} 
\label{fig:degdistkmeans}
\end{figure}

\begin{figure}
\centerline{\includegraphics[width=1.0\columnwidth]{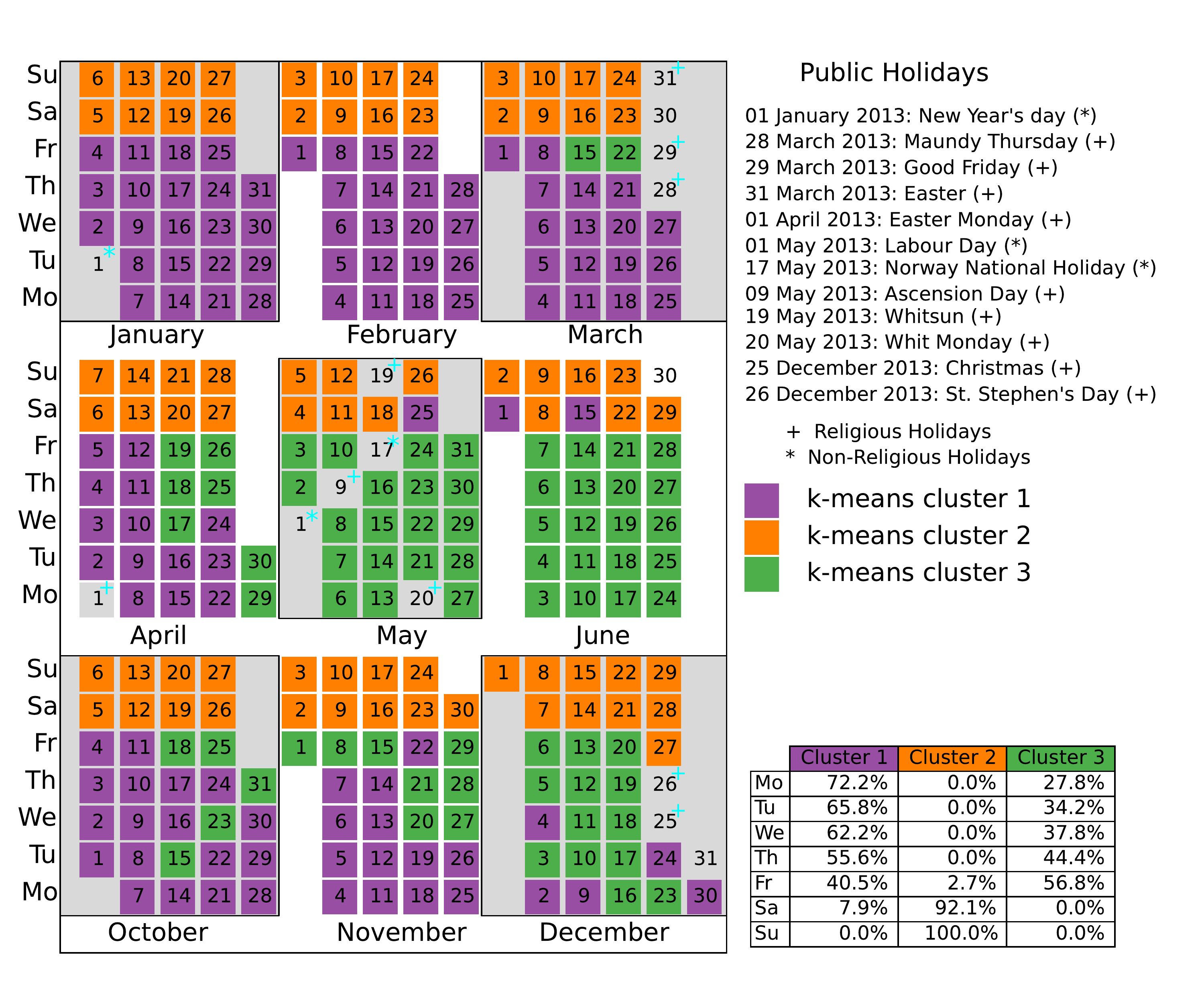}}
\vspace{1pc}
\caption{\textbf{Classification of days from daily call record data using k-means clustering.} Here we used distribution of local clustering of the extracted daily networks as the feature vector. Holidays were removed from the data prior to running the classification routine.} 
\label{fig:clustdistkmeans}
\end{figure}

\begin{figure}
\centerline{\includegraphics[width=1.0\columnwidth]{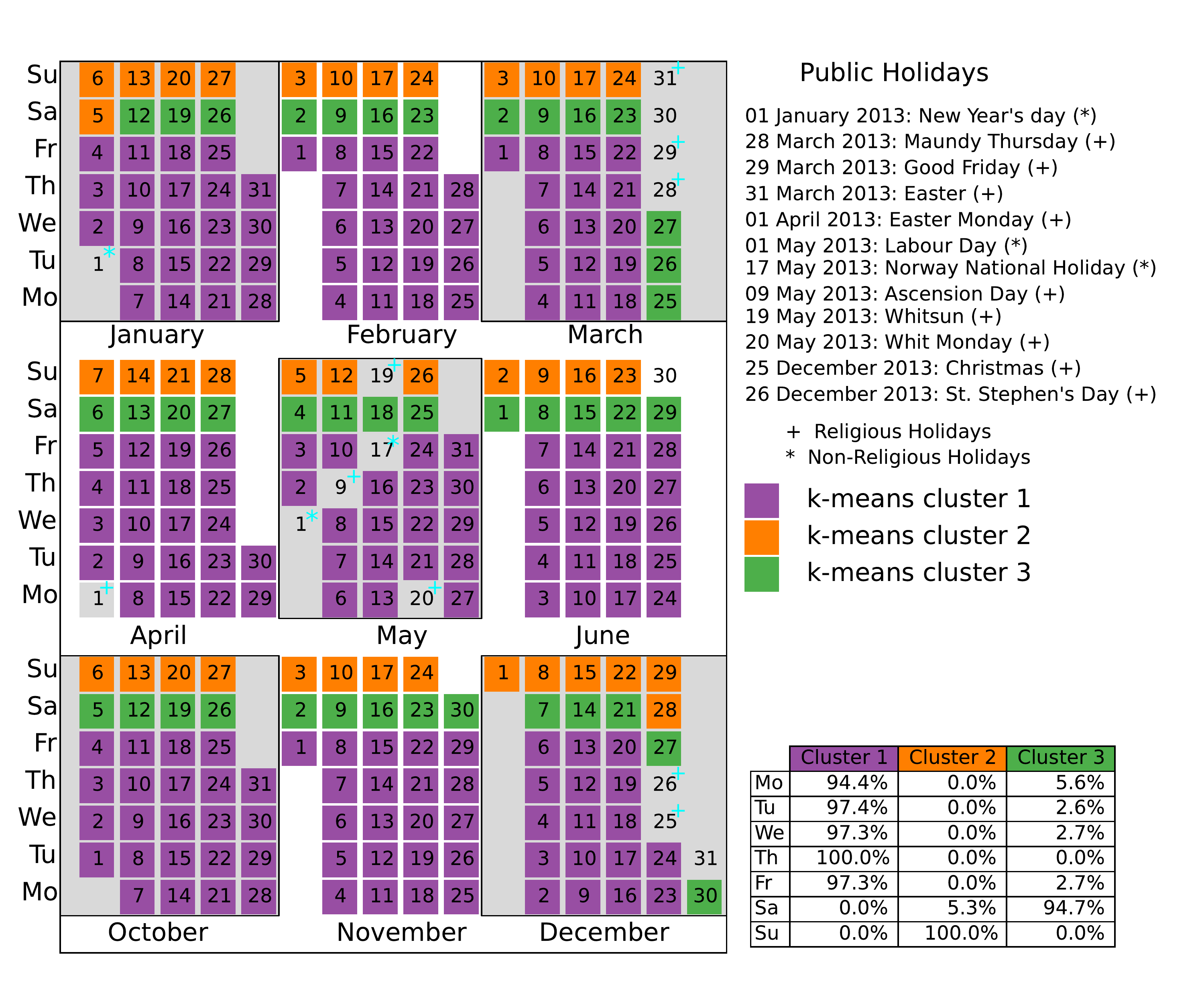}}
\vspace{1pc}
\caption{\textbf{Classification of days from daily call record data using k-means clustering.} The feature vector in this case was composed of features listed in Table~\ref{tab:codeWK}.  Holidays were removed from the data prior to running the classification routine.}
\label{fig:feadistkmeans}
\end{figure}

As shown in Fig.~\ref{fig:degdistkmeans} and  Fig.~\ref{fig:clustdistkmeans}, respectively, we used degree distribution and distribution of local clustering coefficient from daily networks to classify the days into three different clusters. For the distributions, we first generated a histogram of the data and then used a collection binned point probabilities (corresponding to bin heights) as the classification features. In each of these cases, weekdays and weekend days are clearly clustered into different groups.  When using the features from Table~\ref{tab:codeWK},  we observe that k-means is able to distinguish between Saturday and Sunday as well (see Fig.~\ref{fig:feadistkmeans}).

\subsection*{Network sampling}

The accuracies obtained above and in the main text are due, in no small part, to the large quantity of available data. However, in many scenarios the study design does not allow the luxury of the full network from such a massive sample size. In this case, one must use a subsample of the network. To investigate the effect of network subsampling on predictive power, we compare the performance of two subsampling procedures, sampling on geography and snowball sampling. Sampling on geography (via ZIP codes) selects a subset of individuals who live in close proximity to one another, whereas snowball sampling starts with a seed node in the network and branches out from that node following its edges, going several edges away from the seed node, recruiting the nodes along the path to the sample.

We fit the following model:
\begin{equation}\label{invmodel}
\frac{1}{MR_i + \delta} =\beta_0+ \beta_1 X_i + \beta_2 Z_i X_i + \epsilon_i
\end{equation}
where $MR_i$ is the misclassification rate based the $i$th subsample, $X_i$ is the average daily network size (number of nodes) based on the $i$th subsample, $Z_i$ is an indicator for if the $i$th observation was based on a ZIP code subsample, $\delta=0.01$ is a shift used to avoid division by $0$, and $E[\epsilon_i]=0$ with $\mbox{Var}(\epsilon_i)=\sigma_i^2$. In addition, we force $\beta_0=(5/7+\delta)^{-1}$ to reflect the fact that when sample size is 0 the misclassification rate is $5/7$ (corresponding to the classifier that predicts every day to be a weekday). Due to strong heteroskedastic errors in this model and a strong presence of outliers, model \eqref{invmodel} is fit using least-absolute-deviations regression \cite{barrodale1973improved}.  We test for a difference in the misclassification rates when comparing snowball to ZIP code sampling. This test corresponds to the hypotheses $H_0: \beta_2=0$ and $H_A: \beta_2\neq 0$. To perform inference on $\hat{\beta}_2$ we simulate the null distribution of $\hat{\beta}_2$ by permuting sampling-type labels ($Z_i$). Due to ZIP code and snowball samples having different distributions in average network size (ZIP code samples tend to be larger), we put observations in bins of width $20$ and permute $Z_i$ labels only on observations within each bin. This preserves the distribution of average network size amongst both types of sampling procedures.

\begin{figure}
\centerline{\includegraphics[scale=.45]{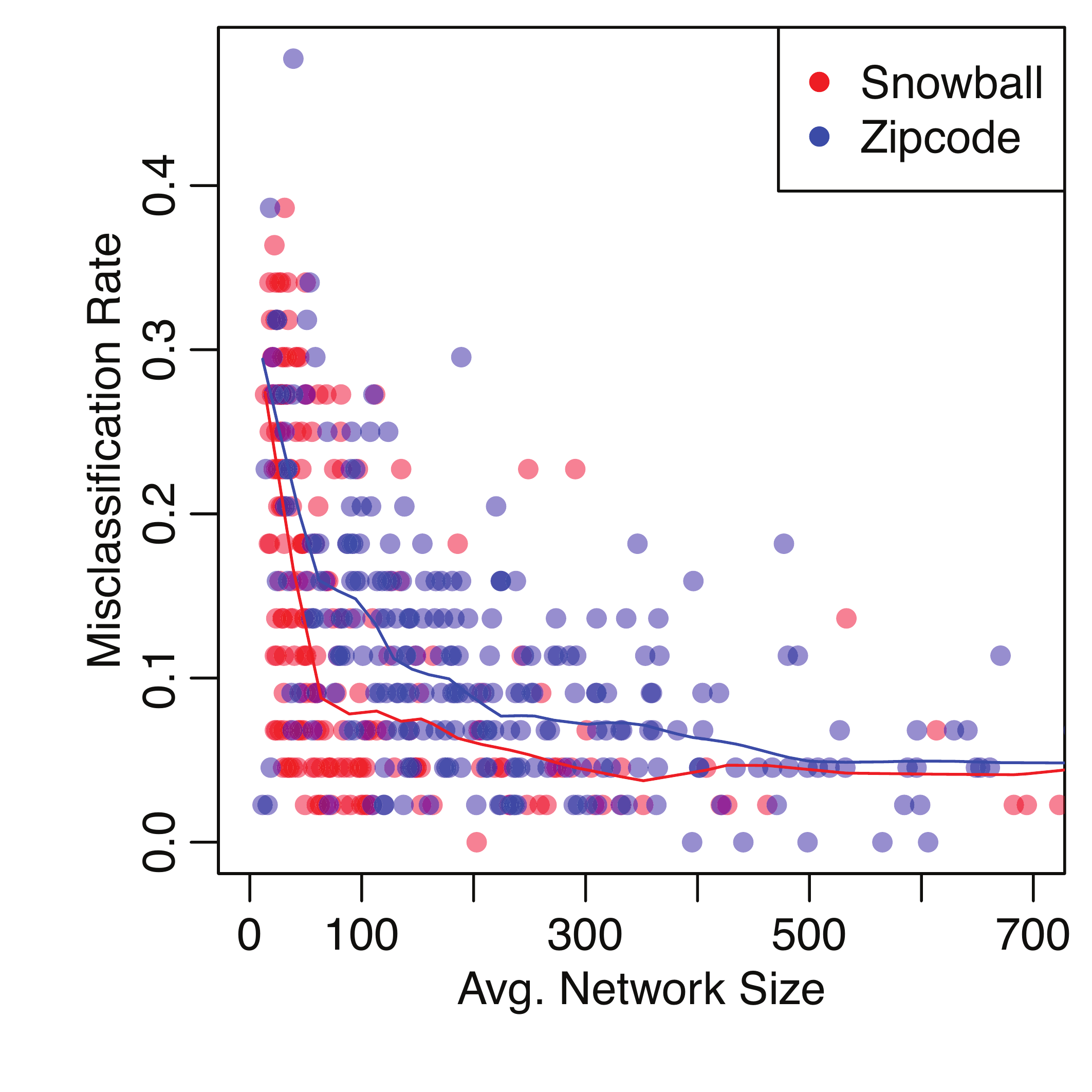}}
\vspace{1pc}
\caption{\textbf{Misclassification rates of classifiers built on network subsamples.} Each point represents a random forest classifier built from a subsample of the full network. Networks are constructed on a daily basis. The inverse-transformed misclassification rate is used to fit the regression model. Snowball samples with radii 4, 5, and 6 are included. Each subsampled network has varying network size on any given day because the networks are constructed using edge lists, so if a node in the original subsampled network has degree 0 on a particular day, then that node will not appear in that day's network. A smooth is fit to each point cloud, one for snowball subsamples and one for ZIP code subsamples.} \label{SubsampleMRs}
\end{figure}

The misclassification rate of the weekend/weekday random forest classifier for all considered snowball and ZIP code subsamples are displayed in Fig.~\ref{SubsampleMRs}. After fitting model \eqref{invmodel}, which relates network size and subsampling procedure to misclassification rate, we found that $\hat{\beta}_2=-0.03$. This implies that when holding the network size fixed, the slope of the expected misclassification rate on the inverse scale is $-0.03$ lower for ZIP code sampling than it is for snowball sampling. This change in slope is significant ($\mbox{p-value}=6.5\cdot 10^{-5}$), implying that, on average, snowball sampling yields networks that have features that inform classification of weekends and weekdays better than those from ZIP code samples of equivalent sample size.

There is a clear tendency for the ZIP code subsamples to have higher misclassification rates relative to the size of the subsample than is the case for snowball subsamples. A large part of the reason for this greater misclassification in ZIP code subsamples could be because the feature measuring the fraction of ties that are within the same ZIP code holds no meaning for ZIP code subsamples (trivially the fraction is always one). As seen in Fig.~3, this feature is vitally important to classification of weekends from weekdays, so ZIP code subsamples suffer without it.

To perform a snowball sample based on a given day's network, a random seed node is selected and all nodes within a distance of $4$ are included in the subsample. This is repeated for each day in the data set as well as for radii of 5 and 6. In each case, if the resulting network has fewer than 50 nodes, a new random seed node is selected until the subsample of sufficient size is acquired. ZIP code subsamples include all individuals from the same ZIP code regardless of their social connections. Each ZIP code in the country matching with at least 50  active customers in the data set was used as a separate subsample. Altogether there were 247 snowball subsamples and 293 ZIP code subsamples included in the analysis. The same binary random forest classification procedure was replicated for each subsample as was performed on the full data set, and the resulting misclassification rates for each subsample were recorded.

\clearpage

\clearpage
\clearpage
\clearpage

\begin{figure}
\centerline{\includegraphics[scale=.55]{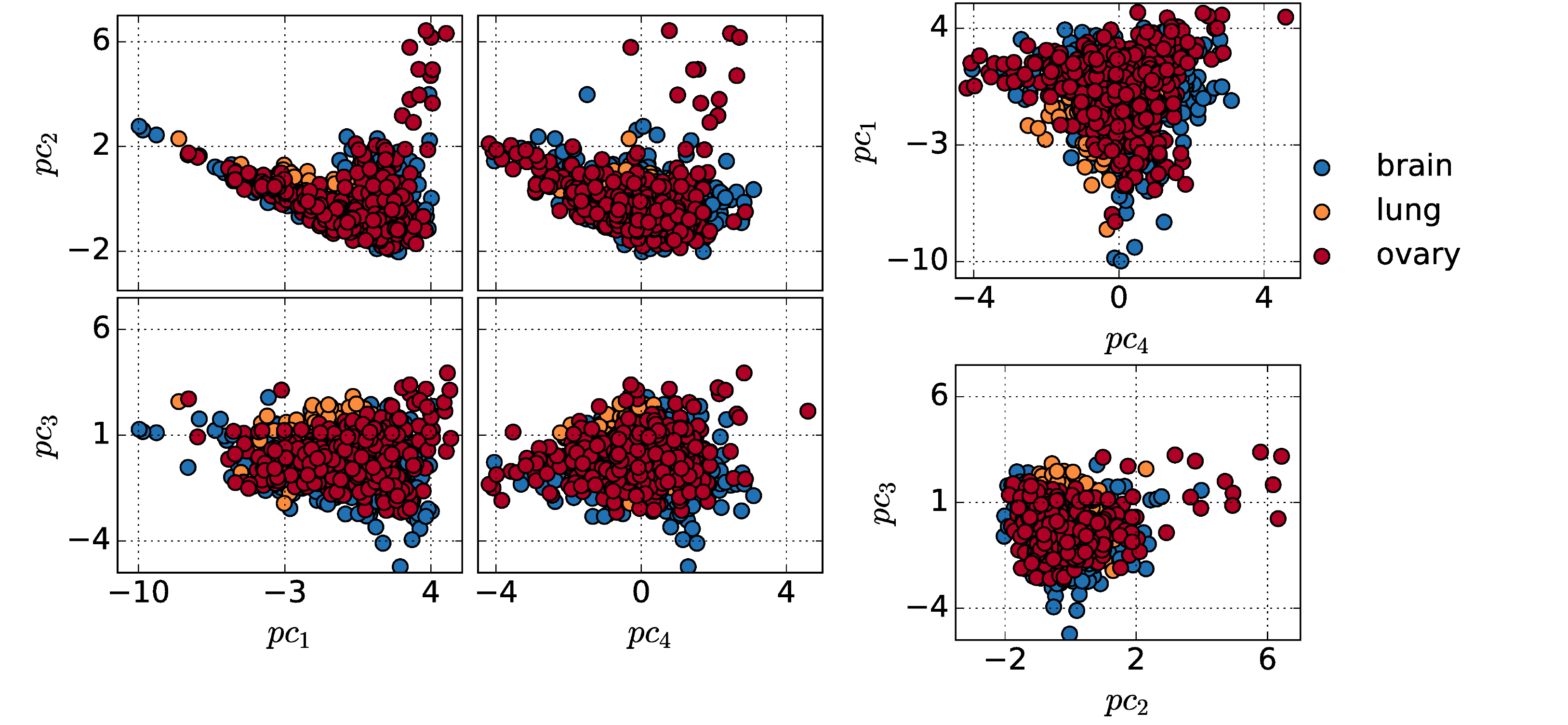}}
\vspace{1pc}
\caption{\textbf{PCA based classification of cancer types.}}
\label{fig:cantyppca}
\end{figure}

\begin{table}[]
\centering
\begin{tabular}{|l|l|l|l|l|l|}
\hline
Data set            & RF         & KNN         & GK          & DGK         & PSCN        \\
\hline
COLLAB             & $76.5\pm1.68$ & $72.69\pm0.80$ & $72.84\pm0.28$ & $73.09\pm0.25$ & $72.60\pm2.15$  \\
\hline
IMDB-BINARY       & $72.4\pm4.69$ & $37.03\pm1.90$  & $65.87\pm0.98$ & $66.96\pm0.56$ & $71.00\pm2.29$    \\
\hline
IMDB-MULTI        & $47.8\pm3.55$ & $42.40\pm2.70$  & $43.89\pm0.38$ & $44.55\pm0.52$ & $45.23\pm2.84$ \\
\hline
REDDIT-BINARY     & $88.7\pm1.99$ & $87.63\pm0.82$ & $77.34\pm0.18$ & $78.04\pm0.39$ & $86.30\pm1.58$  \\
\hline
REDDIT-MULTI-5K  & $50.9\pm2.07$ & $49.04\pm0.77$ & $41.01\pm0.17$ & $41.27\pm0.18$ & $49.10\pm0.70$   \\
\hline
REDDIT-MULTI-12K & $42.7\pm1.28$ & $38.21\pm0.49$ & $31.82\pm0.08$ & $32.22\pm0.10$ & $41.32\pm0.42$ \\
\hline
\end{tabular}
\caption{\textbf{Classification accuracy for online social network data sets.} The classification accuracy ($\%$) for the five different classifiers are compared. 10-fold cross-validation was used to obtain out-of-sample accuracy estimates and their standard errors.}
\label{tab:socnetclass}
\end{table}

\begin{table}
\begin{center}
\begin{tabular}{|r|l|}
\hline
Feature label & Feature name \\
\hline
Gassor	& Degree assortativity in the gene projection network 	\\
\hline
Tassor	&  Degree assortativity in the TF projection network	\\
\hline
Bclus	&  Average bipartite clustering coefficient in the bipartite network 	\\
\hline
Gclus	& 	Average clustering coefficient in the gene projection network\\ 
\hline
Tclus		& Average clustering coefficient in the TF projection network   \\
\hline 
Gtri	&  Number of triangles in the gene projection network	\\
\hline
Ttri	&  Number of triangles in the TF projection network\\
\hline
AvgDeg	&  Average degree in the bipartite network	\\
\hline
Gavgdeg	&  Average degree in the gene projection network	\\ 
\hline
Tavgdeg	&  Average degree in the TF network	\\
\hline
cctM	& Mean closeness centrality in the bipartite network 	\\
\hline
cctV	& 	Variance of closeness centrality in the bipartite network\\
\hline
nrcM	& Mean node redundancy in the bipartite network	\\
\hline
nrcV	&  Variance of node redunancy in the bipartite network	\\
\hline
\end{tabular}
\end{center}
\caption{\textbf{Description of feature labels for tumor type classification.} A more detailed description for the feature labels used in Figure 4 and \ref{fig:cantyppca}. Here TF stands for transcription factor.}
\label{tab:codeBIO}
\end{table}

\clearpage

\begin{table}
\begin{center}
\begin{tabular}{|r|l|}
\hline
Feature label & Feature name \\
\hline
NumNodes	& Number of nodes 	\\
\hline
NumEdges		& Number of edges   \\
\hline 
NumTri	&  Number of triangles	\\
\hline
ClustCoef	&  Global clustering coefficient	\\
\hline
DegAssort	&  Degree assortativity coefficient\cite{newman2003mixing}	\\ 
\hline
AvgDeg	&  Average degree	\\
\hline
FracF	&  Fraction of nodes that are female	\\
\hline
FracMF	&  Fraction of edges that are male-female	\\
\hline
AvgAgeDif	&  Average age difference (absolute value) over edges	\\
\hline
FracSameZip	& 	Fraction of edges that share the same ZIP code\\ 
\hline
DegPC1-4	&  Principal components of degree distribution	\\
\hline
ClusPC1-4	&  Principal components of clustering distribution	\\
\hline
\end{tabular}
\end{center}
\caption{\textbf{Description of feature labels for weekday/weekend classification.} A more detailed description for the feature labels used in Figure 2, 3, \ref{fig:degdistPCA} (c)  and \ref{fig:feadistkmeans}. This includes some redundant features such as NumTri, NumEdges, and DegPC1, all of which are strongly correlated ($\rho > 0.9$) with NumNodes.}
\label{tab:codeWK}
\end{table}

\end{document}